\documentclass{ws-ijmpa}
\usepackage[super,compress]{cite}
\usepackage{graphicx}
\usepackage{hyperref}
\usepackage{color}

\begin{document}
\markboth{F. Ahmed}{Higher-dimensional topologically charged wormhole}

\catchline{}{}{}{}{}

\title{ Higher-dimensional topologically charged traversable defect wormhole with non-exotic matter }

\author{Faizuddin Ahmed\footnote{E-mail: faizuddinahmed15@gmail.com}
}

\address{Department of Physics, University of Science \& Technology Meghalaya, Ri-Bhoi, Meghalaya, 793101, India
}

\maketitle

\begin{history}
\received{Day Month Year}
\revised{Day Month Year}
\end{history}

\begin{abstract}
In this study, we explore a topologically charged higher-dimensional traversable defect wormhole, with a specific emphasis on five dimensions. Particularly noteworthy is the fact that the matter-energy distribution attributed to this charged wormhole configuration adheres to the weak energy condition, thus presenting an instance of a five-dimensional wormhole supported by non-exotic matter. Furthermore, our analysis shows that scalar quantities related to space-time curvature and parameters associated with the matter-energy distribution remain finite at the wormhole throat. Moreover, they diminish as distance extends toward infinity, indicating the asymptotic flatness inherent in our model.


\end{abstract}

\ccode{PACS numbers: 04.50.-h; 04.50.Kd; 14.80.Hv}

\section{Introduction}

Wormholes are hypothetical tunnels that connect two regions of the same universe or even distinct universes through a narrow throat. The concept of traversable wormholes was groundbreaking when Morris and Thorne introduced their seminal work on it \cite{MT,MT2}. Unlike previously considered wormholes, such as the Einstein-Rosen bridge \cite{ER} (an analysis of Einstein and Rosen's work can be related to Flamm's earlier work \cite{LF1}, with English translations available in Refs. \cite{LF2,LF3}), or Wheeler's microscopic charge-carrying wormholes \cite{JAW}, traversable wormholes are defined in a way that allows for the two-way travel of objects. However, an exact solution for a wormhole was first provided by Ellis \cite{HGE}, with a similar solution independently proposed by Bronnikov \cite{KAB} in the same year. These wormholes, collectively known as the Ellis-Bronnikov wormhole, have been extensively explored in the scientific literature. Additionally, a time-independent spherically symmetric solution by T. Kodama \cite{TK} was known in the literature.

Morris-Thorne wormholes particularly introduced a novel perspective on the possibility of time travel within the framework of general relativity theory. The origins of this concept can be traced back to 1949 when K. Gödel presented his seminal work on a rotating universe \cite{GO}. Numerous wormhole models have been constructed and thoroughly investigated, significantly contributing to our understanding of these objects. Some of these models include Visser's wormhole models \cite{MV}, a self-consistent wormhole \cite{EGH}, rotating wormholes \cite{Teo}, static and dynamic plane-symmetric wormholes \cite{jpsl}, Morris-Thorne wormholes with a cosmological constant \cite{jpsl2}, a general class of spherically symmetric wormholes \cite{AD}, varying cosmological constants wormholes \cite{FR}, stationary and cylindrically symmetric rotating wormholes without exotic matter \cite{KB,KB2}, solutions to the Einstein–Maxwell equations for rotating cylindrical symmetry wormholes \cite{KB3}, NUT traversable wormholes \cite{GC}, and static, spherically symmetric wormhole solutions with a minimally coupled scalar field \cite{KB4}. In higher-dimensional theories, wormhole solutions have been constructed in various settings, including in five-dimensional theory \cite{gd}, in six-dimensional vacuum space-time \cite{avm}, in $n$-dimensional theory \cite{TT}, with compact dimensions \cite{AD3}, one-parameter extension of the Schwarzschild black hole \cite{ASMV}, regular Fisner space-time \cite{gg1}, extension of Reissner-Nordström space-time \cite{gg2}, and globally regular black bounce space-times \cite{fsnl0}.

A primary concern regarding wormhole space-times is their potential violation of one or more energy conditions, which are essential principles in physics. This violation raises questions about whether the energy-momentum tensor associated with these models corresponds to physically realistic matter, though it's not inherently impossible. The consensus is that any physical system must adhere to the weak and null energy conditions \cite{SWH}. In the case of traversable wormholes, exotic matter threads the wormhole throat, generating repulsion against its collapse. These wormholes lack an event horizon, allowing observers to move freely across the universe. Research has shown that a black hole solution with horizons can be transformed into a wormhole solution by introducing exotic matter, thereby stabilizing the wormhole \cite{JPS}. Traversable wormhole solutions must satisfy the flare-out condition to maintain their geometry and keep the throat open. The presence of exotic matter leads to violations of the null energy condition (NEC) and weak energy condition (WEC), essential properties for traversable wormholes. The violation of the NEC, the weakest energy condition, results in violations of the WEC and strong energy conditions (SEC).

Theoretical investigation of these exotic spacetime geometries continues to enhance our understanding of general relativity and other fundamental aspects of physics. Remarkable progress has been made in the field of wormhole physics, addressing the challenge of exotic matter. It has been recognized that vacuum solutions of the field equations naturally satisfy the energy conditions. Building on this research, a recent study was reported in \cite{FRK}, where a traversable defect wormhole was discussed that does not require exotic matter and satisfies both the weak and null energy conditions. In \cite{FRK2}, a vacuum defect traversable wormhole was constructed, representing a significant advancement in this area in higher dimensions. Furthermore, in \cite{FRK3}, a novel type of traversable wormhole solution was proposed, eliminating the need for exotic matter, marking a significant step forward in making wormholes more feasible. In this direction, a Schwarzschild-Klinkhamer traversable wormhole was introduced in \cite{FA2}, incorporating a cosmic string and global monopole while maintaining compliance with the energy conditions. This development highlights the potential interplay between different physical phenomena in the context of wormholes. Additionally, in \cite{FRK4}, a Schwarzschild-like defect wormhole, enriching our understanding of the various types of traversable wormholes, was presented. Extending this wormhole research into higher dimensions, a five-dimensional vacuum-defect wormhole was presented in \cite{FRK5,FA}, and a Schwarzschild-like five-dimensional topologically charged wormhole in \cite{FRK6}. This advancement opens up new possibilities for exploring wormholes in broader contexts.

It is widely believed that topological defects formed through a spontaneous symmetry-breaking mechanism in the early universe \cite{kk2,kk3,kk4,kk6}. One specific example of these topological defects is the global monopole, a spherically symmetric entity arising from the self-coupling triplet of scalar fields $\phi^a$ ($a=1,2,3$). In \cite{MBAV}, the authors presented an approximate solution of Einstein's field equations for a metric in the region outside the core of the GM, which forms due to global symmetry breaking. Neglecting the mass term, the monopole metric describes a space with a deficit solid angle \cite{ERBM}. When the GM mass term is not negligible, the solution describes a kind of Schwarzschild black hole with an additional GM charge \cite{nd}. Moreover, in \cite{Shi}, an exact solution to the nonlinear equation describing a global monopole in flat space was presented. It has been demonstrated in \cite{DPP} that the gravitational field of GM can lead to the clustering of matter and Cosmic Microwave Background (CMB) anisotropies. Furthermore, GM may play a role in seeding the formation of supermassive black holes \cite{RB}. Studies have indicated that this type of topological defects induces a negative gravitational potential, resulting in a repulsive gravitational field \cite{Shi, kk12}. The same topic in the asymptotically dS/AdS scenario was explored in \cite{nd2}, which found that the gravitational potential can be either repulsive or attractive, depending on the value of the cosmological constant. In \cite{kk4}, the potential role of topological defects in our Universe has been systematically expounded. Numerous researchers have illustrated the crucial role of topological defects in cosmological structure formation and the evolution of the cosmos \cite{MY,kk13,kk14,kk15,mm1,mm2,mm3,mm4,kk5,mm5,mm6}.

In five-dimensional relativity theory, two prevailing formulations, namely membrane theory and induced-matter (or space-time-matter) theory, have taken center stage in research areas for a long time. The former framework introduces an exponential factor related to the extra coordinate and its impact on four-dimensional space-time, while the latter introduces a quadratic factor (see Ref. \cite{ss1} for detailed discussions). Numerous solutions within the realm of five-dimensional theory have been formulated in the literature (see Refs. \cite{ss11,ss12,ss15,ss14,ss16,ss17,ss23,ss24}).

In four-dimensional relativity theory, charged wormhole space-times with or without exotic matter distribution has been studied by several authors in the literature (see, Refs. \cite{FA2,FA3,FA4} and related references there in). In the context of five-dimensional theory, a comprehensive vacuum-defect wormhole space-time was presented in \cite{FA}. This 5D wormhole model in an extension of vacuum-defect four-dimensional wormhole given by Klinkhamer in Ref. \cite{FRK2} (also see \cite{FRK}). This extension represents a noteworthy advancement in the wormhole structures within the broader multi-dimensional context. The metric describing this five-dimensional vacuum-defect wormhole is given by \cite{FA} $(c=1=\hbar=G)$
\begin{equation}
    ds^2=-dt^2+\Big(1+\frac{b^2}{\xi^2}\Big)^{-1}\,d\xi^2+(\xi^2+b^2)\,(d\theta^2+\sin^2 \theta\,d\phi^2+\cos^2 \theta\,d\chi^2).\label{K1}
\end{equation}
Here, the coordinate $\chi \in (0, \infty)$ represents the extra dimension, characterized spatially, and $t \in (-\infty, +\infty)$, $\xi \in (-\infty, +\infty)$, $\theta \in [0, \pi)$, and $\phi \in [0, 2\,\pi)$.

In $n$ dimensions, general relativity can be described by the Einstein–Hilbert action with a Lagrangian $\mathcal{L}_{matter}$ of matter fields appearing in the theory
\begin{equation}
    S=\int\,d^nx\,\Big(\frac{1}{\kappa}\,R+\mathcal{L}_{matter}\Big)\,\sqrt{-g},\quad \kappa=\frac{8\,\pi\,G}{c^4}.\label{K2}
\end{equation}
Using geometrized units $G=c=8\,\pi=1$, the variation of the action (\ref{K2}) with respect to the metric tensor leads to the Einstein field equations
\begin{equation}
    G_{MN}=R_{MN}-\frac{1}{2}\,g_{MN}\,R=T_{MN},\label{K3}
\end{equation}
where the energy–momentum tensor $T_{MN}$ is given by the variation of the matter field Lagrangian $\mathcal{L}_{matter}$ as follows:
\begin{equation}
    T_{MN}=\frac{2}{\sqrt{-g}}\,\frac{\delta (\sqrt{-g}\,\mathcal{L}_{matter})}{\delta g^{MN}}\,.\label{K4}
\end{equation}
From the trace of (\ref{K3}), one may express the Ricci scalar $R$ as
\begin{equation}
    R=-\frac{2}{n-2}\,T,\quad\quad T=T^{M}_{\,\,M}\,.\label{K5}
\end{equation}
For the five-dimensional geometry, we have
\begin{equation}
    R=-\frac{2}{3}\,T\,.\label{K6}
\end{equation}
Therefore, the field equations in five-dimensions can be written as
\begin{equation}
    G_{AB}=R_{AB}-\frac{1}{2}\,g_{AB}\,R=T_{AB}\quad \mbox{or}\quad R_{AB}=T_{AB}-\frac{1}{3}\,g_{AB}\,T,\quad (A,B=0,1,..,4)\,.\label{K7}
\end{equation}

In this article, we aim to examine a five-dimensional defect traversable wormhole characterized by a topological charge of global monopoles, a generalized version of the vacuum-defect five-dimensional wormhole presented in (\ref{K1}). The primary objective of this study revolves around presenting the fundamental features and properties inherent in this higher-dimensional wormhole and analyze the influence of global monopoles. Furthermore, the matter-energy distribution associated with this wormhole is also investigated and demonstrate the weak energy condition.

\section{Analysis of five-dimensional topologically charged traversable defect wormhole}

In this section, we extend our investigation to a generalized version of the five-dimensional vacuum-defect wormhole (\ref{K1}), incorporating a global monopole charge. This generalized traversable wormhole presents a non-vacuum solution of the field equations, featuring non-exotic matter. Hence, we commence this section by introducing a modified version of the five-dimensional wormhole metric (\ref{K1}), termed as the topologically charged defect wormhole, utilizing the following line-element ansatz in the chart $(t, \xi, \theta, \phi, \chi)$:
\begin{eqnarray}
ds^2=-dt^2+\Bigg(1+\frac{b^2}{\xi^2}\Bigg)^{-1}\,\frac{d\xi^2}{\alpha^2}+(\xi^2+\lambda^2)\,\Big(d\theta^2+\sin^2 \theta\,d\phi^2+\cos^2 \theta\,d\chi^2\Big),\label{1}
\end{eqnarray}
where $\lambda, b$ are positive constants with $b>\lambda$, and $0 < \alpha=(1-8\,\pi\,G\,\eta^2_{0})<1$ characterizes the global monopole parameter. Here $G$ being the universal constant of gravitation and $\eta_0$ the dimensionless volumetric mass density of the pointlike GM \cite{ERBM}. The introduction of GM parameter $\alpha$ into the above line-element significantly alters the physical and geometrical properties of the space-time. This parameter $\alpha$ is introduced analogously to the Schwarzschild-like five-dimensional topologically charged wormhole \cite{FRK6} and the four-dimensional topologically charged wormholes, such as the Klinkhamer-type defect wormhole \cite{FA2}, the generalized Schwarzschild-Simpson-Visser-type wormhole \cite{FA3}, the rotating wormhole \cite{FA4}, and the Ellis-Bronnikov-type wormhole \cite{EPJC, EPL}. Moreover, gravitational field of global monopole within curved space-times has been discussed in general relativity and modified gravity (see, for examples, \cite{MBAV,ERBM,HSR1,HSR2,HSR3,JRN}). 

It can be observed that in the limit $\alpha \to 1$ and $\lambda^2=b^2$, space-time (\ref{1}) yields a five-dimensional vacuum-defect wormhole configuration \ref{K1}, extensively discussed in Ref. \cite{FA}. Additionally, at the surface $\chi=0$, the metric (\ref{1}) leads back to a four-dimensional topologically charged Klinkhamer-type defect wormhole, as extensively elaborated in Ref. \cite{FA2}. Moreover, in the limit $\alpha \to 1$ and $\chi=0$ surface, the metric (\ref{1}) retrieves a four-dimensional defect wormhole devoid of exotic matter, as extensively studied in Ref. \cite{FRK}.

This five-dimensional metric (\ref{1}) is a non-vacuum solution of the field equations with nonzero components of the Einstein tensor $G^{A}_{\,B}$, where $A, B=0,1,2,3,4$ are given by
\begin{eqnarray}
&&G^{t}_{t}=-\frac{3\,(1-\alpha^2)}{\xi^2+\lambda^2}=-R/2,\nonumber\\
&&G^{\xi}_{\xi}=\frac{3\,\Big(b^2\,\alpha^2-\lambda^2+\xi^2\,(-1+\alpha^2)\Big)}{(\xi^2+\lambda^2)^2},\nonumber\\
&&G^{\theta}_{\theta}=\frac{\Big(-b^2\,\alpha^2+\xi^2\,(-1+\alpha^2)+\lambda^2\,(-1+2\,\alpha^2)\Big)}{(\xi^2+\lambda^2)^2}=G^{\phi}_{\phi}=G^{\chi}_{\chi},\label{2}
\end{eqnarray}
where $R=g_{AB}\,R^{AB}$ is the Ricci scalar. 

The different scalar quantities associated the space-time curvature, such as the Kretschmann scalar curvature, $\mathcal{K}=R^{ABCD}\,R_{ABCD}$, the Ricci scalar, $R=g_{AB}\,R^{AB}$, and the quadratic Ricci invariant $\mathcal{R}=R^{AB}\,R_{AB}$ for the metric (\ref{1}) are given by
\begin{eqnarray}
\mathcal{K}&=&\frac{12\,\Big[(\xi^2+\lambda^2)^2-2\,(\xi^2+\lambda^2)(\xi^2+b^2)\,\alpha^2+(\lambda^4-2\,\lambda^2\,b^2+2\,b^4+2\,\xi^2\,b^2+\xi^4)\Big]}{(\xi^2+\lambda^2)^4},\nonumber\\
R&=&\frac{6\,(1-\alpha^2)}{\xi^2+\lambda^2},\nonumber\\
\mathcal{R}&=&\frac{3\,\Big[3\,(\lambda^2-b^2)^2\,\alpha^4+\Big\{-2\,(\xi^2+\lambda^2)+(\lambda^2+b^2+2\,\xi^2)\,\alpha^2\Big\}^2\Big]}{(\xi^2+\lambda^2)^4}.
\label{3}
\end{eqnarray}
Also the nonzero components of the Riemann curvature tensor $R^{A}_{BCD}$ are given by 
\begin{eqnarray}
    &&R^{\xi}_{\theta\theta\xi}=\frac{(\lambda^2-b^2)\,\alpha^2}{\xi^2+\lambda^2},\nonumber\\
    &&R^{\xi}_{\phi\phi\xi}=\frac{(\lambda^2-b^2)}{\xi^2+\lambda^2}\,\alpha^2\,\sin^2 \theta,\nonumber\\
    &&R^{\xi}_{\chi\chi\xi}=\frac{(\lambda^2-b^2)}{\xi^2+\lambda^2}\,\alpha^2\,\cos^2 \theta,\nonumber\\
    &&R^{\theta}_{\xi\theta\xi}=\frac{(-\lambda^2+b^2)\,\xi^2}{(\xi^2+\lambda^2)^2(\xi^2+b^2)}=R^{\phi}_{\xi\phi\xi}=R^{\chi}_{\xi\chi\xi},\nonumber\\
    &&R^{\phi}_{\theta\phi\theta}=\Bigg(1-\frac{\xi^2+b^2}{\xi^2+\lambda^2}\,\alpha^2\Bigg)=R^{\chi}_{\theta\chi\theta},\nonumber\\ 
    &&R^{\theta}_{\phi\phi\theta}=-R^{\phi}_{\theta\phi\theta}\,\sin^2 \theta=R^{\chi}_{\phi\chi\phi},\nonumber\\ 
    &&R^{\theta}_{\chi\chi\theta}=-R^{\phi}_{\theta\phi\theta}\,\cos^2 \theta=R^{\phi}_{\chi\chi\phi}.
    \label{5}
\end{eqnarray}

At the wormhole throat, $\xi=0$, these scalar quantities are finite given by 
\begin{eqnarray}
   &&\mathcal{K}|_{\xi=0}=\frac{24\,\Big(\lambda^4-\lambda^2\,b^2\,\alpha^2-\lambda^2\,b^2+b^4\Big)}{\lambda^8},\nonumber\\
   &&R|_{\xi=0}=\frac{6\,(1-\alpha^2)}{\lambda^2},\nonumber\\
   &&\mathcal{R}|_{\xi=0}=\frac{3\,\Big[3\,(\lambda^2-b^2)^2\,\alpha^4+\Big\{-2\,\lambda^2+(\lambda^2+b^2)\,\alpha^2\Big\}^2\Big]}{\lambda^8}.
   \label{8}
\end{eqnarray}
Also the Riemann curvature tensor components (\ref{5}) in this case ($\xi=0$) becomes
\begin{eqnarray}
    &&R^{\xi}_{\theta\theta\xi}|_{\xi=0}=\frac{(\lambda^2-b^2)\,\alpha^2}{\lambda^2},\nonumber\\
    &&R^{\xi}_{\phi\phi\xi}|_{\xi=0}=\frac{(\lambda^2-b^2)}{\lambda^2}\,\alpha^2\,\sin^2 \theta,\nonumber\\ 
    &&R^{\xi}_{\chi\chi\xi}|_{\xi=0}=\frac{(\lambda^2-b^2)}{\lambda^2}\,\alpha^2\,\cos^2 \theta,\nonumber\\
    &&R^{\theta}_{\phi\phi\theta}|_{\xi=0}=\Big(-1+\frac{b^2}{\lambda^2}\,\alpha^2 \Big)\,\sin^2 \theta=R^{\chi}_{\phi\chi\phi}|_{\xi=0},\nonumber\\
    &&R^{\phi}_{\theta\phi\theta}|_{\xi=0}=\Big(1-\frac{b^2}{\lambda^2}\,\alpha^2\Big)=R^{\chi}_{\theta\chi\theta}|_{\xi=0},\nonumber\\ 
    &&R^{\theta}_{\chi\chi\theta}|_{\xi=0}=\Big(-1+\frac{b^2}{\lambda^2}\,\alpha^2 \Big)\,\cos^2 \theta=R^{\phi}_{\chi\chi\phi}|_{\xi=0}.
    \label{10}
\end{eqnarray}
From the above analysis, one can see that different scalar quantities associated with the space-time curvature of the wormhole (\ref{1}) are finite at the wormhole throat $\xi=0$ and vanishes for $\xi \to \pm\,\infty$. Thus, the five-dimensional topologically charged traversable defect wormhole given by the line-element (\ref{1}) is free-from curvature singularity. 

As the wormhole space-time (\ref{1}) possesses the nonzero Einstein's tensor $G^{AB}$ given by Eq. (\ref{2}), we choose the energy-momentum tensor to be an anisotropic fluid as the source, taking the following form:
\begin{equation}
    T^{A}_{\,\,\,B}=\mbox{diag}(-\rho, p_{\xi}, p_{t}, p_{t}, p_{\chi}),\quad\quad T=T^{A}_{\,\,A}=-\rho+p_{\xi}+2\,p_{t}+p_{\chi},
    \label{energy}
\end{equation}

where $\rho$ represents the energy-density, and $p_{\xi}, p_{t}, p_{\chi}$ represents the pressure components along $\xi$-direction, tangential directions, and along $\chi$-direction, respectively. 

Therefore, from the field equation $G^{A}_{\,\,\,B}=T^{A}_{\,\,\,B}$ and using equations (\ref{2}) and (\ref{energy}), we obtain different physical quantities as follows: 
\begin{eqnarray}
    &&\rho=\frac{3\,(1-\alpha^2)}{\xi^2+\lambda^2}=\frac{R}{2}>0,\nonumber\\
    &&p_{\xi}=\frac{3}{(\xi^2+\lambda^2)^2}\,\Big[b^2\,\alpha^2-\lambda^2+\xi^2\,(-1+\alpha^2)\Big],\nonumber\\ 
    &&p_{t}=\frac{\Big[-b^2\,\alpha^2+\xi^2\,(-1+\alpha^2)+\lambda^2\,(-1+2\,\alpha^2)\Big]}{(\xi^2+\lambda^2)^2}=p_{\chi}.
    \label{6}
\end{eqnarray}

Since the global monopole parameter lies in the ranges $0 < \alpha < 1$ in the gravitation and cosmology, we can see that the energy-density of the matter-energy distribution is always positive, thus, satisfying the weak energy condition. Noted that by substituting Eq. (\ref{6}) into the expression $T$ in Eq. (\ref{energy}), one can easily verify the relation (\ref{K6}) for the given Ricci scalar $R$ in Eq. (\ref{3}). Thus, the five-dimensional wormhole space-time (\ref{1}) satisfied the Einstein field equations (\ref{K7}) with ansiotropic fluid as matter content.

At the wormhole throat, $\xi=0$, these physical parameters (\ref{6}) becomes finite given by
\begin{eqnarray}
    &&\rho|_{\xi=0}=\frac{3\,(1-\alpha^2)}{\lambda^2}>0,\nonumber\\
    &&p_{\xi}|_{\xi=0}=\frac{3\,(b^2\,\alpha^2-\lambda^2)}{\lambda^4},\nonumber\\
    &&p_{t}|_{\xi=0}=\frac{-b^2\,\alpha^2+\lambda^2\,(-1+2\,\alpha^2)}{\lambda^4}.
    \label{7}
\end{eqnarray}
One can easily show that at $\xi \to \pm\,\infty$, the energy-density as well as pressure components vanishes. 

\vspace{0.2cm}
\begin{center}
    {\bf Special Case Corresponds to $\lambda^2=b^2$}
\end{center}
\vspace{0.2cm}

In this part, we consider a special case corresponds to $\lambda^2=b^2$. In that case, the five-dimensional line-element (\ref{1}) can be written as
\begin{eqnarray}
    ds^2=-dt^2+\Big(1+\frac{b^2}{\xi^2}\Big)^{-1}\,\frac{d\xi^2}{\alpha^2}+(\xi^2+b^2)\,\Big(d\theta^2+\sin^2 \theta\,d\phi^2+\cos^2 \theta\,d\chi^2\Big).
    \label{a1}
\end{eqnarray}
For this space-time (\ref{a1}), the nonzero components of the Einstein tensor $G^{A}_{\,\,\,B}$ are given by
\begin{eqnarray}
    &&G^{t}_{t}=-R/2=G^{\xi}_{\xi},\nonumber\\
    &&G^{\theta}_{\theta}=-R/6=G^{\phi}_{\phi}=G^{\chi}_{\chi},
    \label{a2}
\end{eqnarray}
where $R$ is the Ricci scalar. This Ricci scalar along with the Kretschmann scalar curvature, and the Ricci invariant for the metric (\ref{a1}) are given by
\begin{eqnarray}
   &&R=\frac{6\,(1-\alpha^2)}{\xi^2+b^2},\nonumber\\
   &&\mathcal{K}=\frac{12\,(-1+\alpha^2)^2}{(\xi^2+b^2)^2},\nonumber\\
   &&\mathcal{R}=\frac{12\,(-1+\alpha^2)^2}{(\xi^2+b^2)^2}.
   \label{a3}
\end{eqnarray}
These scalar quantities associated the space-time curvature are finite at $\xi=0$ and vanishes for $\xi \to \pm\,\infty$.

The nonzero components of the Riemann curvature tensor are as follows:
\begin{eqnarray}
    &&R^{\theta}_{\phi\phi\theta}=(-1+\alpha^2)\,\sin^2 \theta=-R^{\chi}_{\phi\chi\phi},\nonumber\\
    &&R^{\theta}_{\chi\chi\theta}=(-1+\alpha^2)\,\cos^2 \theta=R^{\phi}_{\chi\chi\phi},\nonumber\\ 
    &&R^{\phi}_{\theta\phi\theta}=1-\alpha^2=R^{\chi}_{\theta\chi\theta}.
    \label{a5}
\end{eqnarray}
From equations (\ref{a2})--(\ref{a5}), it is clear that in the limit $\alpha \to 1$, all these physical entities vanish. In that case, the five-dimensional wormhole metric (\ref{a1}) becomes a vacuum-defect solution (\ref{K1}) obtained in Ref. \cite{FA}.

In this special cases, we considering the same stress-energy tensor (\ref{energy}) as done earlier. Therefore, from the field equations $G^{A}_{\,\,\,B}=T^{A}_{\,\,\,B}$ and using equations (\ref{energy}) and (\ref{a2}), we obtain different physical quantities as follows: we obtain
\begin{eqnarray}
    &&\rho=\frac{3\,(1-\alpha^2)}{\xi^2+b^2},\nonumber\\
    &&p_{\xi}=-\frac{3\,(1-\alpha^2)}{\xi^2+b^2},\nonumber\\
    &&p_{t}=\frac{(-1+\alpha^2)}{\xi^2+b^2}=p_{\chi}.
    \label{a6}
\end{eqnarray}

Since the parameter $0 < \alpha < 1$ in the gravitation and cosmology, thus, we find the following conditions \cite{SWH}
\begin{eqnarray}
    WEC&:&\quad \rho>0,\nonumber\\
    WEC_{t}&:&\quad \rho+p_{t}>0,\nonumber\\
    WEC_{\xi}&:&\quad \rho+p_{\xi}=0,\nonumber\\
    WEC_{\chi}&:&\quad \rho+p_{\chi}>0.
    \label{a7}
\end{eqnarray}
From above it is clear that the matter-energy distribution satisfies the weak energy condition (WEC).

At the wormhole throat, $\xi=0$, the physical quantities given in Eq. (\ref{a6}) are finite given by
\begin{eqnarray}
    &&\rho|_{\xi=0}=\frac{3\,(1-\alpha^2)}{b^2}>0,\nonumber\\
    &&p_{\xi}|_{\xi=0}=-\frac{3\,(1-\alpha^2)}{b^2},\nonumber\\
    &&p_{t}|_{\xi=0}=\frac{(-1+\alpha^2)}{b^2}.
    \label{a8}
\end{eqnarray}
In this special case, we see that the topological charge of global monopole parameter $\alpha$ control the geometrical and physical properties of the space-time geometry under consideration.

\section{Conclusions}

One of the fascinating aspects of general relativity is its allowance for hypothetical geometries with nontrivial topological structures. Misner and Wheeler \cite{MW} introduced these features of space-time as solutions to the Einstein field equations, termed wormholes. A ``wormhole" comprises a tunnel with two ends, offering a shortcut between distant regions of the universe. Despite some lacking observational evidence, wormholes are often conceptualized as part of the black hole (BH) family \cite{SAH}. The simplest example is the Schwarzschild wormhole, linking different parts of the universe through a bridge. However, this type of wormhole isn't traversable since it doesn't facilitate two-way communication between distant regions, leading to the contraction of the wormhole throat.

Numerous wormhole models have been proposed in the literature, each showcasing intriguing properties. Some of these models adhere to the weak energy condition, eliminating the need for exotic matter-energy distributions. Our contribution in this study represents a significant advancement in the understanding of higher-dimensional wormholes, building upon recent proposals for five-dimensional wormhole models. The incorporation of topological charge from a global monopole profoundly alters the characteristics of the space-time geometry, a subject we have meticulously explored. Importantly, our findings affirmed that the five-dimensional wormhole model complies with the weak energy condition, thus requiring non-exotic matter-energy distributions. Our analysis also shows that scalar quantities intricately linked to space-time curvature, along with the physical parameters governing matter-energy distribution, exhibit bounded behavior as $\xi \to 0$ and tend towards zero as $\xi \to \pm\,\infty$, highlighting the pervasive regularity inherent in our charged traversable defect wormhole configuration. Several investigations into the stability of wormhole geometries, both in lower dimensions and in four dimensions, have been conducted under radial perturbations. Moving forward, we aim to explore the stability of the wormhole geometry under consideration in the future work.

\section*{Acknowledgments}

We would like to thank the anonymous referee for their valuable comments and helpful suggestions. F.A. acknowledges the Inter-University Centre for Astronomy and Astrophysics (IUCAA), Pune, India, for granting a visiting associateship.


\begin{thebibliography}{0}

\bibitem{MT} M. S. Morris and K. S. Thorne, \href{https://doi.org/10.1119/1.15620}{Am. J. Phys. {\bf 56}, 395 (1988)}.

\bibitem{MT2} 
M. Morris, K. S. Thorne and U. Yurtsever, \href{https://doi.org/10.1103/PhysRevLett.61.1446}{Phys. Rev. Lett. {\bf 61}, 1446 (1988).}

\bibitem{ER} 
A. Einstein and N. Rosen, \href{https://doi.org/10.1103/PhysRev.48.73}{Phys. Rev. {\bf 48}, 73 (1935)}.

\bibitem{LF1}
L. Flamm, Zeit. Phys. {\bf XVII}, 448 (1916).

\bibitem{LF2}
G. W. Gibbons, \href{https://doi.org/10.1007/s10714-015-1907-3}{Gen. Relativ. Gravit. {\bf 47}, 72 (2015).}

\bibitem{LF3}
G. W. Gibbons, \href{https://doi.org/10.1007/s10714-015-1907-3}{Gen. Relativ. Gravit. {\bf 47}, 71 (2015).}

\bibitem{JAW} J. A. Wheeler, {\it Geometrodynamics}, Academic press, New York (1962).

\bibitem{HGE} 
H. G. Ellis, \href{https://doi.org/10.1063/1.1666161}{J. Math. Phys. {\bf 14}, 104 (1973)}.

\bibitem{KAB} 
K. A. Bronnikov, \href{https://www.actaphys.uj.edu.pl/R/4/3/251}{Acta Phys. Pol. B {\bf 4}, 251 (1973)}.

\bibitem{TK} 
T. Kodama, \href{https://doi.org/10.1103/PhysRevD.18.3529}{Phys. Rev. D 18, 3529 (1978).}

\bibitem{GO}
K. G\"{o}del, \href{https://doi.org/10.1103/RevModPhys.21.447}{Rev. Mod. Phys. {\bf 21}, 447 (1949).}

\bibitem{MV}
M. Visser, \href{https://doi.org/10.1103/PhysRevD.39.3182}{Phys. Rev. {\bf D 39}, 3182(R) (1989).}

\bibitem{EGH}
E. G. Harris, \href{https://doi.org/10.1119/1.17310}{Am. J. Phys. {\bf 61}, 1140 (1993).}

\bibitem{Teo}
E. Teo, \href{https://doi.org/10.1103/PhysRevD.58.024014}{Phys. Rev. {\bf D 58}, 024014 (1998).}

\bibitem{jpsl}
J. P. S. Lemos and F. S. N. Lobo, \href{https://doi.org/10.1103/PhysRevD.78.044030}{Phys. Rev. {\bf D 78}, 044030 (2008).}


\bibitem{jpsl2} 
J. P. S. Lemos and F. S. N. Lobo, S. Q. de Oliveira, \href{https://doi.org/10.1103/PhysRevD.68.064004}{Phys. Rev. {\bf D 68}, 064004 (2003).}

\bibitem{AD} 
A. DeBenedictis and A. Das, \href{https://doi.org/10.1088/0264-9381/18/7/304}{Class. Quantum Grav. {\bf 18}, 1187 (2001).}

\bibitem{FR}
F. Rahaman, M. Kalam, M. Sarker, A. Ghosh and B. Raychaudhuri, \href{https://doi.org/10.1007/s10714-006-0380-4}{Gen. Relativ. Gravit. {\bf 39}, 145 (2007).}

\bibitem{KB}
K. A. Bronnikov, V.  G. Krechet and J. P. S. Lemos, \href{https://doi.org/10.1103/PhysRevD.87.084060}{Phys. Rev. {\bf D 87}, 084060 (2013)}.

\bibitem{KB2} 
K. A. Bronnikov and V. G. Krechet, \href{https://doi.org/10.1142/S0217751X16410220}{Int. J. Mod. Phys. {\bf A 31}, 1641022 (2016)}.

\bibitem{KB3} 
K. A. Bronnikov, V. G. Krechet and V. B. Oshurko, \href{https://doi.org/10.3390/sym12081306}{Symmetry {\bf 12} (8), 1306 (2020)}.

\bibitem{GC}
G. Clément, D. Gal'tsov and M. Guenouche, \href{https://doi.org/10.1103/PhysRevD.93.024048}{Phys. Rev. {\bf D 93}, 024048 (2016).}

\bibitem{KB4} 
K. A. Bronnikov and S. V. Sushkov, \href{https://doi.org/10.1088/0264-9381/27/9/095022}{Class. Quantum Grav. {\bf 27}, 095022 (2010).}

\bibitem{gd}
G. Dotti, J. Oliva and R. Troncoso, \href{https://doi.org/10.1103/PhysRevD.75.024002}{Phys. Rev. {\bf D 75}, 024002 (2007).}

\bibitem{avm}
A. V. Aminova and P. I. Chumarov, \href{https://doi.org/10.48550/arXiv.1307.0316}{arXiv:1307.0316 [gr-qc].}

\bibitem{TT}
T. Torii and H. Shinkai, \href{https://doi.org/10.1103/PhysRevD.88.064027}{Phys. Rev. {\bf D 88}, 064027 (2013).}

\bibitem{AD3}
A. DeBenedictis and A. Das, \href{https://doi.org/10.1016/S0550-3213(03)00051-8}{Nucl. Phys. {\bf B 653}, 279 (2003).}

\bibitem{ASMV} 
A. Simpson and M. Visser, \href{https://doi.org/10.1088/1475-7516/2019/02/042}{JCAP 02 ({\bf 2019}) 042}.

\bibitem{gg1} 
K. A. Bronnikov, \href{https://doi.org/10.1103/PhysRevD.106.064029}{Phys. Rev. {\bf D 106}, 064029 (2022)}.

\bibitem{gg2} 
K. A. Bronnikov and R. K. Walia, \href{https://doi.org/10.1103/PhysRevD.105.044039}{Phys. Rev. {\bf D 15}, 044039 (2022)}.

\bibitem{fsnl0} 
F. S. N. Lobo, M. E. Rodrigues, M. V. d. S. Silva, A. Simpson and M. Visser, \href{https://doi.org/10.1103/PhysRevD.103.084052}{Phys. Rev. {\bf D 103}, 084052 (2021)}.

\bibitem{SWH}
S. W. Hawking and G. F. R. Ellis, {\it The Large Scale Structure of Space-Time}, Cambridge University Press, Cambridge (1973).

\bibitem{JPS}
J. P. Lemos, \href{https://doi.org/10.1016/0370-2693(95)00533-Q}{Phys. Lett. B {\bf 353}, 46 (1995).}

\bibitem{FRK} 
F. R. Klinkhamer, \href{https://doi.org/10.5506/APhysPolB.54.5-A3}{Acta Phys. Polon. B {\bf 54}, 5-A3 (2023)}.

\bibitem{FRK2} 
F. R. Klinkhamer,  \href{https://doi.org/10.5506/APhysPolB.54.7-A3}{Acta Phys. Pol. B {\bf 54}, 7-A3 (2023)}.

\bibitem{FRK3} 
F. R. Klinkhamer, \href{https://doi.org/10.55318/bgjp.2024.51.1.042}{Bulg. J. Phys. {\bf 51} (1), 042 (2024)}.

\bibitem{FA2}
F. Ahmed, \href{https://doi.org/10.1088/1475-7516/2023/11/010}{JCAP 11 ({\bf 2023}) 010.}

\bibitem{FRK4} 
Z.-L. Wang, \href{https://doi.org/10.48550/arXiv.2307.01678}{arXiv: 2307.01678 [gr-qc]}.

\bibitem{FRK5} 
F. R. Klinkhamer, \href{https://doi.org/10.48550/arXiv.2307.12876}{arXiv: 2307.12876 [gr-qc]}.

\bibitem{FA} 
F. Ahmed, \href{https://doi.org/10.48550/arXiv.2308.11938}{arXiv: 2308.11938 [gr-qc].}

\bibitem{FRK6}
F. Ahmed, \href{https://doi.org/10.5506/APhysPolB.55.1-A3}{Acta Phys. Polo. B {\bf 55}, 1-A3 (2024).}

\bibitem{kk2}
T. W. B. Kibble, \href{https://doi.org/10.1016/0370-1573(80)90091-5}{Phys. Rept. {\bf 67}, 183 (1980).}

\bibitem{kk3}
T. W. B. Kibble, \href{https://www.actaphys.uj.edu.pl/R/13/10/723}{Acta Phys. Pol. {\bf B 13}, 723 (1982).}

\bibitem{kk4}
A. Vilenkin and E. P. S. Shellard, {\it Cosmic Strings and Other Topological Defects}, Cambridge University Press, Cambridge (1994). 

\bibitem{kk6}
M. Hindmarsh, A.-C. Davis and R. Brandenberger, \href{https://doi.org/10.1103/PhysRevD.49.1944}{Phys. Rev. {\bf D 49}, 1944 (1994). }

\bibitem{MBAV}
M. Barriola and A. Vilenkin, \href{https://doi.org/10.1103/PhysRevLett.63.341}{Phys. Rev. Lett. {\bf 63}, 341 (1989).}

\bibitem{ERBM}
E. R. Bezerra de Mello, \href{https://doi.org/10.1590/S0103-97332001000200012}{Braz. J Phys. {\bf 31}, 211 (2001).}

\bibitem{nd}
N. Dadhich, K. Narayan and U. A. Yajnik, \href{https://doi.org/10.1007/BF02845552}{Pramana-J. Phys. {\bf 50}, 307 (1998).}

\bibitem{Shi}
X. Shi and X. Li, \href{https://doi.org/10.1088/0264-9381/8/4/019}{Class. Quantum Grav. {\bf 8}, 761 (1991).} 

\bibitem{DPP}
D. P. Bennett and S. H. Rhie, \href{https://doi.org/10.1103/PhysRevLett.65.1709}{Phys. Rev. Lett. {\bf 65}, 1709 (1990).}

\bibitem{RB}
R. Brandenberger and H. Jiao, \href{https://doi.org/10.1088/1475-7516/2020/02/002}{JCAP 02 ({\bf 2020}) 002.}

\bibitem{kk12}
D. Harari and C. Loust’o, \href{https://doi.org/10.1103/PhysRevD.42.2626}{Phys. Rev. {\bf D 42}, 2626 (1990).}

\bibitem{nd2}
X.-Z. Li and J.-G Hao, \href{https://doi.org/10.1103/PhysRevD.66.107701}{Phys. Rev. D 66, 107701 (2002).}

\bibitem{MY}
M. Yamaguchi, \href{https://doi.org/10.1103/PhysRevD.64.081301}{Phys. Rev. {\bf D 64}, 081301(R) (2001).}

\bibitem{kk13}
A. Vilenkin, \href{https://doi.org/10.1103/PhysRevLett.72.3137}{Phys. Rev. Lett. {\bf 72}, 3137 (1994).}

\bibitem{kk14}
R. Basu, A. H. Guth and A. Vilenkin, \href{https://doi.org/10.1103/PhysRevD.44.340}{Phys. Rev. {\bf D 44}, 340 (1991).}

\bibitem{kk15}
R. Basu and A. Vilenkin, \href{https://doi.org/10.1103/PhysRevD.50.7150}{Phys. Rev. {\bf D 50}, 7150 (1994).}

\bibitem{mm1}
R. H. Brandenberger, \href{https://doi.org/10.1007/BF02827491}{Pramana-J Phys. {\bf 51}, 191 (1998).}

\bibitem{mm2} 
A. Gangui, {\it Lecture Notes for the First Bolivian School on Cosmology}, La Paz, 24-28 September, 2001, pp. 78 [\href{https://doi.org/10.48550/arXiv.astro-ph/0110285}{arXiv:astro-ph/0110285.}]

\bibitem{mm3}
R. Durrer, \href{https://doi.org/10.1016/S1387-6473(99)00008-1}{New Astron. Rev. {\bf 43}, 111 (1999)}.

\bibitem{mm4}
M. Sakellariadou, {\it Invited lectures in the NATO ASI / COSLAB (ESF) School ``Patterns of symmetry breaking'', September 2002 (Cracow), pp. 28} [\href{https://doi.org/10.48550/arXiv.hep-ph/0212365}{arXiv:hep-ph/0212365}]. 

\bibitem{kk5}
R. Brandenberger, \href{https://doi.org/10.1142/S0217751X9400090X}{Int. J. Mod. Phys. {\bf A 9}, 2117 (1994).}

\bibitem{mm5}
R. Durrer and Z. -H. Zhou, \href{https://doi.org/10.1103/PhysRevD.53.5394}{Phys. Rev. {\bf D 53}, 5394 (1996).}

\bibitem{mm6}
R. Durrer, M. Kunz and A. Melchiorri, \href{https://doi.org/10.1016/S0370-1573(02)00014-5}{Phys. Rept. {\bf 364}, 1 (2002).}

\bibitem{ss1}
P. S. Wesson, \href{https://doi.org/10.1142/6029}{\it Five-Dimensional Physics}, Singapore: World Scientific (2006).

\bibitem{ss11}
V. Dzhunushaliev and V. Folomeev, \href{https://doi.org/10.1142/S0217732314500254}{Mod. Phys. Lett. {\bf A 29}, 1450025 (2014).}

\bibitem{ss12}
T. Liko and P. S. Wesson, \href{https://doi.org/10.1063/1.1926168}{J. Math. Phys. {\bf 46}, 062504 (2005).}

\bibitem{ss15}
J. B. Fonseca-Neto, C. Romero and F. Dahia, \href{https://doi.org/10.1142/S0217732306019001}{Mod. Phys. Lett. {\bf A 21}, 525 (2006).}

\bibitem{ss14}
P. S. Wesson, H. Liu and S. S. Seahra, \href{https://doi.org/10.48550/arXiv.gr-qc/0003012}{Astron. Astrophys. {\bf 358}, 425 (2000).}

\bibitem{ss16}
T. Fukui, S S. Seahra and P. S Wesson, \href{https://doi.org/10.1063/1.1407836}{J. Math. Phys. {\bf 42}, 5196 (2001).}

\bibitem{ss17}
C.-M. Chen, \href{https://doi.org/10.1088/0264-9381/18/20/302}{Class. Quantum Grav. {\bf 18}, 4179 (2001).}

\bibitem{ss23}
P. S Wesson, \href{https://doi.org/10.1088/0264-9381/19/11/306}{Class. Quantum Grav. {\bf 19}, 2825 (2002).}

\bibitem{ss24}
S. S. Seahra and P. S. Wesson, \href{https://doi.org/10.1063/1.1623617}{J. Math. Phys. {\bf 44}, 5664 (2003).}

\bibitem{FA3}
F. Ahmed, \href{https://doi.org/10.1088/1475-7516/2023/11/082}{JCAP 11 ({\bf 2023}) 082.}

\bibitem{FA4}
F. Ahmed, \href{https://doi.org/10.5506/APhysPolB.54.11-A3}{Acta Phys. Pol. B {\bf 54}, 11-A3 (2023).}

\bibitem{EPJC}
H. Aounallah, A. R. Soares and R. L. L. Vitória, \href{https://doi.org/10.1140/epjc/s10052-020-7980-0}{Eur. Phys. J. C {\bf 80}, 447 (2020).} 

\bibitem{EPL}
F. Ahmed, \href{https://doi.org/10.1209/0295-5075/acbb21}{EPL  {\bf 141}, 54001 (2023).} 

\bibitem{HSR1}
R. D. Lambaga and H. S. Ramadhan, \href{https://doi.org/10.1140/epjc/s10052-018-5906-x}{Eur. Phys. J. C {\bf 78}, 436 (2018).}

\bibitem{HSR2}
I. Prasetyo and H. S. Ramadhan, \href{https://doi.org/10.1007/s10714-015-1998-x}{Gen. Relativ. Gravit. {\bf 48}, 10 (2016)}.

\bibitem{HSR3}
A. M. Kusuma, B. N. Jayawiguna, and H. S. Ramadhan, \href{https://doi.org/10.1103/PhysRevD.104.124045}{Phys. Rev. {\bf D 104}, 124045 (2021).}

\bibitem{JRN}
J. R. Nascimento, G. J. Olmo, A. Yu. Petrov, P. J. Porfirio and A. R. Soares, \href{https://doi.org/10.1103/PhysRevD.99.064053}{Phys. Rev. {\bf D 99}, 064053 (2019).}

\bibitem{MW}
C. W. Misner and J. A. Wheeler, \href{https://doi.org/10.1016/0003-4916(57)90049-0}{Ann. Phys. (NY) {\bf 2}, 525 (1957).}

\bibitem{SAH}
S. A. Hayward, \href{https://doi.org/10.48550/arXiv.gr-qc/0203051}{arXiv:gr-qc/0203051 [gr-qc].}


\end{thebibliography}
\end{document}